# The Yule distribution and frailty—

# a note on spurious preferential attachment

Birgitte Freiesleben de Blasio[1], Odd O. Aalen[1],

[1]Department of Biostatistics, Institute of Basic Medical Sciences, University of Oslo, PO Box 1122, Blindern, N-0317 Oslo, Norway

Corresponding author:

Dr. Birgitte Freiesleben de Blasio
Department of Biostatistics
Institute of Basic Medical Science
University of Oslo
b.f.d.blasio@medisin.uio.no
phone :0047-22 85 15 08
fax: 0047-22 85 13 13

# ABSTRACT

Preferential attachment is a popular generative mechanism to explain the widespread observation of power law distributed networks. We introduce an alternative explanation for the phenomenon by allowing the link growth rates to vary across the nodes according to a randomized Poisson process. The distribution of rates, which reproduces the degree distribution of a preferential attachment process (Yule process) is derived analytically. We demonstrate with use of simulations that the degree distribution and growth rates in single time intervals are similar for the random process and the preferential attachment process. Structural differences are analyzed by examining the joint degree distribution and network coreness.

# INTRODUCTION

In recent years, empirical data on large-scale networks have become available and provided valuable insights into the structure of real-world connected systems. Many complex networks are found to have a power law distribution of links.

In an influential paper Barabasi & Albert [1] explained the emergence of power laws graphs by a preferential attachment (PA) process, where new nodes link to existing nodes with a probability $p_i$ that is proportional to the connectivity of the nodes, with $p_i = k_i / \sum_j k_j$. The great appeal of the PA model is that is transcends simple data fitting and offers a mechanistic interpretation of the observed phenomenon. Today, most of existing models aimed at reproducing power law graphs incorporate a PA mechanism of some kind; numerous examples can be found in [2].

However, already in 1925 Yule [3] had published a closely related stochastic branching model. The Yule process is general and has been adapted to networks [4,5,6].

In this article we will focus on an alternative explanation for the formation of power law graphs: The distribution could arise if one assumes each node to have a different probability of attracting a new link. In this case—unlike the PA scenario—the power law graph does not arise because the nodes display diversifying behavior during the growth process. The variation is a constant and an inherent property of the nodes.

Frailty is a general term used in event history analyses to describe unobserved heterogeneity in data. We shall derive analytically the frailty distribution of a randomized Poisson graph, which reproduces the Yule distribution. Indeed, random graphs with arbitrary degree distributions like the configuration model [7] have been studied extensively; though the main interest has been devoted to properties of static graphs of infinite size. The fitness point of view has also been considered by other authors, e.g. [8].

Here we focus on the Yule process, which appears not to have been studied. The Yule process has the advantage of being a general preferential attachment mechanism where its randomized Poisson graph counter part is analytically tractable. This gives the possibility for direct comparison of the evolving Yule graph and the frailty variant through numerical simulations.

Despite the popularity of the PA concept, the interpretation of real-world graphs in terms of the mechanism(s) responsible for creating them is not trivial. First, the degree distribution provides a first order description of the network structure at a given point in time. It does not supply information on how it was formed. Hence, model fit to an empirical distribution cannot be used as indicator for the relevance of a given generative process. Second, the argument in support of PA is often derived from the finding of cumulative advantage in sociological systems. Here individuals with early success tend to be selected at the expense of people who at the outset are less fortunate [9]. For example, a paper with many citations is likely to be cited again because it will tend to pop up during a literature search.

However, in some types of networks this "popularity is attractive" scheme is not obvious; e.g. in a sexual network [10] where the connectivity of nodes is not displayed, and therefore cannot be used directly as a selection criterion. One could also argue that scientific citation networks should reflect to a larger degree the true importance of a paper (inherent criteria), and to a minor degree the crowding around already popular papers. Naturally, the two mechanisms may coexist, and together with other factors act to shape the network topology.

The point we want to raise is that the Yule process and a random process acting on a heterogeneous sample may give a good fit of the same heavy-tailed distribution, although their interpretation of the causal mechanism is completely opposite. It is interesting to note that a similar controversy surrounds the related Polya process [11]. The Polya distribution can be derived for true 'contagion' where the occurrence of an event increases the probability that a new event will take place [12]. In contrast, as shown by

Greenwood and Yule [13], the same distribution may arise from a stratified Poisson process, hence the name 'spurious contagion'.

How do we know if PA takes place in evolving networks?

It has been suggested that correlation between link growth rates and node degree can be used to distinguish the PA process, e.g. with use of graphical methods [14,15]. The idea is to measure the intensity by which groups of nodes with identical connectivity $[k]$ acquire new links during a small time interval $\Delta t$. By plotting the mean number of new partners $<\Delta k>$ during $\Delta t$ as function of $[k]$, PA will manifest itself as a linear dependence.

As will be shown, this procedure fails to distinguish an evolving frailty graph form the true PA process. In either case the graph has the appearance of PA. The finding suggests that caution should be exercised when interpreting complex network data.

Principally, it is possible to separate the two processes by plotting the growth rates in several disjoint time intervals. In practice, however, given the notoriously noisy data, graphical methods are not advisable. Instead, more advanced topological measures, such as the joint degree distribution may be used to discriminate the processes. Alternatively, in cases where this information is not available, sophisticated statistical methods are needed, e.g. with use of a maximum likelihood estimation procedure, comparing the growth rates in consecutive time intervals [16].

# THE YULE DISTRIBUTION

The Yule distribution has its origin in a mutational evolutionary model [3,17,18]. It has been adapted to networks in different ways; here we follow the derivation by Newman [4].

Consider a growing network where new nodes arrive consecutively. At the point of arrival, new nodes make $j_0$ links, where $j_0 = 0,1,2...$ to the existing nodes. If $j_0 = 0$, the model requires an additional 'attractiveness' parameter $a > 0$ in order for the new nodes to engage in the PA process. In between the arrival of new nodes, a total of $m$ links are formed between the existing nodes. Here the target ends are chosen according to their present connectivity $j_i$ with $p_i = j_i / \sum_k j_k$. In principle, at time $t = 0$ the system starts with two nodes connected by $m$ links, so that at time $t = t'$ there are $(m + j_0)t'$ links in a system of size $N = (2 + t')$ nodes. It can be shown with use of a master equation technique [4] that the asymptotic distribution for the number of nodes growing to infinity has the form:

$$p_Y(J = j) = \frac{B(j + a, q)}{B(j_0 + a, q - 1)} = (q - 1)\frac{\Gamma(j + a)\Gamma(j_0 + a + q - 1)}{\Gamma(j_0 + a)\Gamma(j + a + q)} \qquad N \to \infty \qquad (1)$$

where $q = 2 + (j_0 + a)/m$ is the scaling constant:

$$p_Y(J = j) \sim \frac{\Gamma(j + a)}{\Gamma(j + a + q)} \propto j^{-q} \qquad j \to \infty \qquad (2)$$

The Barabasi & Albert PA model, where new nodes enter and link to $m$ different existing nodes chosen with linear preference, arise as a special case of $m = a; j_0 = 0$ and has the scaling property $p(j) \propto j^{-3}$.

## *THE GENERATING FUNCTION*

We aim to find the corresponding frailty probability density function (pdf), which will reproduce the Yule probability distribution function (2).

For this purpose we introduce a random frailty variable $Z$ that is chosen independently for each node. The frailty variable, which is not observable, describes the heterogeneity among the nodes in their tendency to make new links. The distribution of $Z$ on the graph is $f(z)$.

Now, consider a random graph with $N$ nodes, and let $N \to \infty$. The graph has an arbitrary degree distribution described by $P(j)$ on the non-negative integers $j = 0, 1, \ldots$ Assume that given a rate $Z$ it is Poisson distributed with this rate.

Then the following simple relation exists between $P(j)$ and the Laplace transform $L_F$ of the rate distribution $f(z)$:

$$P(J=j) = E\left(\frac{Z^j \exp(-Z)}{j!}\right) = \frac{(-1)^j L_F^{(j)}(1)}{j!} \qquad j = 0, 1, \ldots \tag{3}$$

The probability generating function $G(s)$ of (3) has a power series representation with values of $P(j)$ as the coefficients. With use of (3) we find:

$$G(s) = \sum_{j=0}^{\infty} P(j) s^j = \sum_{j=0}^{\infty} \frac{(-1)^j L_F^{(j)}(1)}{j!} s^j = L_F(1-s) \tag{4}$$

Hence, the probability generating function of a Poisson random distribution with rate parameter $Z$ has the simple relation to the Laplace transform of the rate distribution:

$$G(s) = L_F(1-s) \tag{5}$$

Now, we set the random probability distribution function equal to the Yule distribution, $P(j) = P_Y(j)$. Then we can recover the rate distribution from the inverse Laplace transform $L_F^{-1}(1-s)$.

Inserting (1) we find:

$$G_Y(s) = (q-1) \sum_{k=0}^{\infty} \frac{\Gamma(k+k_0+a)\Gamma(k_0+a+q-1)}{\Gamma(k_0+a)\Gamma(k+k_0+a+q)} s^k$$
$$= \frac{q-1}{k_0+a+q-1} {}_2F_1(1, k_0+a, k_0+a+q; s) \tag{6}$$

where ${}_2F_1$ is the Gauss hypergeometric function, see 15.1.1 of Abramowitz and Stegun[19], henceforward referred to as AS. Notice that the Yule pmf has been re-parameterized $j \to k+k_0$ in order for the summation to start at $k=0$. This gives the following expression for the Laplace transform:

$$L_F(s) = G_Y(1-s) = \frac{q-1}{p+q-1} {}_2F_1(1, p, p+q; 1-s) \tag{7}$$

where we have substituted $p = k_0 + a$. The constant $p$ is identical to the initial proportionality factor for new nodes in the Yule process. The relevant $(p, q)$-values used in the simulations are $q \in [2;3]$ and $p \in ]0;1]$.

## *THE FRAILTY DISTRIBUTION*

The Laplace transform (7) may be inverted using results from queuing theory by Abate & Whitt [20,21]. The resulting frailty pdf has the form:

$$f(z) = \frac{q-1}{p+q-1} \int_0^1 \frac{1}{y} \exp\left(\frac{z(y-1)}{y}\right) B(p, q; y) dy \tag{8}$$

The pdf is a Beta Mixture of Exponentials (BME) averaged with respect to the standard beta distribution $B(p, q; y)$:

$$B(p,q;y) = \frac{\Gamma(p+q)}{\Gamma(p)\Gamma(q)} y^{p-1}(1-y)^{q-1}, \quad 0 \leq y \leq 1, \quad p,q > 0 \tag{9}$$

By a substitution of the variable $x = (1-y)/y$, we can rewrite the integral (8) in the form:

$$f(z) = \frac{1}{B(p,q-1)} \int_0^\infty \exp(-zx)\, x^{q-1}(1+x)^{-(p+q-1)} dx \tag{10}$$

It follows from 13.2.5 of AS that the integral has the solution:

$$f(z) = \frac{(q-1)\Gamma(p+q-1)}{\Gamma(p)} U(q, 2-p, z) \tag{11}$$

where $U(.)$ is the Tricomi hypergeometric function, also known as the confluent hypergeometric function of the second kind.

Alternatively, the function can be written as a combination of regularized confluent hypergeometric functions of the first kind $_1\tilde{F}_1(.)$ (Kummer's function):

$$U(q, 2-p; z) = \pi \csc(\pi p) \left[ \frac{_1\tilde{F}_1(q, 2-p; z)}{\Gamma(q+p-1)} - z^{p-1} \frac{_1\tilde{F}_1(q+p-1, p; z)}{\Gamma(q)} \right] \tag{12}$$

Each frailty is drawn independently from the present distribution.

## SCALING OF THE FRAILTY DISTRIBUTION

The asymptotic behavior of $f(z)$ for $z \to \infty$ can be obtained from Tauberian theory by studying the behaviour of its Laplace transform at $s \to 0$. From properties of the Gauss hypergeometric function [22] it follows that $L_F(s)$ is

of the limiting form $L_F(s) \sim s^{q-1}$. Using Theorem II in [23] we find the scaling property of $f(z)$:

$$f(z) \sim z^{-q} \tag{13}$$

The result is in line with the general finding that combinations of exponentials like (8) result in a power law distribution. This has been studied by Miller [24] in the context of frequencies of words in a text, and by Reed and Hughes [25] who consider processes of exponential growth, which have exponentially distributed survival times.

## *FINITE SIZE SYSTEMS*

The rate frailty model (11) will approach the Yule distribution asymptotically only. However, empirical networks are finite-size networks and hence, the power law scaling will exist in a limited regime. To compare the growth processes, we need to approximate the frailty distribution of a Yule process that evolves during a finite time frame—or equivalently—a Yule distributed network with a finite number of nodes $N$.

In his original paper, Yule generalized the process to a situation with a finite time horizon [3,18]. He showed that the distribution in this case is found by replacing the beta functions in (1) by incomplete beta functions. Hence:

$$P_Y^*(K=k) = \frac{B_\theta(k+p,q)}{B_\theta(p,q-1)} = \frac{1}{B_\theta(p,q-1)} \int_0^\theta t^{k+p-1}(1-t)^{q-1} dt \quad 0 < \theta < 1 \tag{14}$$

We use the notation $P_Y^*(k)$ to designate the time limited Yule distribution. The truncation has the effect of introducing an exponential cut-off on the scaling at $k_{cut} \sim 1/(1-\theta)$ [2]. Thus, we approximate the Yule probability distribution function as:

$$p_Y^*(K=k) \approx CB(k+p,q)\exp(-(1-\theta)k) \tag{15}$$

with a normalization constant:

$$\frac{1}{C} = \sum_{k=0}^{\infty} \exp(-(1-\theta)k) \frac{\Gamma(k+p)\Gamma(q)}{\Gamma(k+p+q)} \quad (16)$$
$$= B(p,q) \, _2F_1(1,p,p+q,\exp(-(1-\theta)))$$

The generating function of (15) is given by:

$$G^*(s) = L_F^*(1-s) \approx C \sum_{k=0}^{\infty} \left( \exp(-(1-\theta)k) \frac{\Gamma(k+p)\Gamma(q)}{\Gamma(k+p+q)} \right) s^k$$
$$= C\, B(p,q) \, _2F_1(1,p,p+q,\exp(-(1-\theta)s)) \quad (17)$$
$$\approx \frac{_2F_1(1,p,p+q,\theta s)}{_2F_1(1,p,p+q,\theta)} \simeq \frac{q-1}{(p+q-1)\theta} \, _2F_1(1,p,p+q,\theta s)$$

In the last equation we have used that for $\theta \to 1$ $\exp(-(1-\theta)) \approx \theta + O[1-\theta]^2$, and $B(p,q) \, _2F_1(1,p,p+q,\theta) \approx B(p,q-1)\theta$. The result (17) implies that the scale on the Yule process is carried over to the frailty variable.

Using (8) and (10), the frailty distribution is given by:

$$f^*(z) \approx \frac{q-1}{(p+q-1)\,\theta} \int_0^1 \frac{1}{y} \exp\left( \frac{z\theta(y-1)}{y} \right) B(p,q;y) dy$$
$$= \frac{\theta \exp(-(1-\theta)z)}{B(p,q-1)} \int_0^{\infty} \exp(-z\theta x) x^{q-1} (1+x)^{-(p+q-1)} dx \quad (18)$$

A simple way to generate the distribution (18) is by making a random variate $A$ from the beta distribution (9) with parameters $p$ and $q-1$. Then the variate $A$ is transformed into $B = A/(1-A)$. Another random exponentially distributed variate $C$ with mean $\mu = 1/\theta$ is generated [21]. The variable $D = BC$ will have the correct form, since:

$$P(D > Z) = \int_x P(C > Zx) P(Y = x) = \int_0^{\infty} \exp(-z\theta x) x^{q-1} (1+x)^{-(p+q-1)} dx \quad (19)$$

in accordance with the integral in (18). What remains is to scale the variate $W$ with a factor $\exp(-(1-\theta)W)$, where the scale $\theta$ is chosen based on the size $N$ and the scale factor $q$ of the Yule network of interest.

## *SIMULATION MODEL*

In the simulations we start with a small number of nodes $N_0 = 20$. The nodes are randomly connected in a way so that each node has exactly $k_0$ link at start of the simulations. The graphs grow with an average of $m$ links for each new node that enters, and each new node makes $k_0$ links to existing nodes at arrival. The procedure runs the following way:

(i) **Yule graph:**
A new links is generated with probability $p_{link} = m/(m+1)$. The origin of the link is chosen randomly among the nodes. The target node is chosen proportional to the present number of links, $p_i = k_i / \sum_j k_j$. With probability $p_{node} = 1/(m+1)$ a new node arrives. New nodes enter with either zero or one link. In the first case the nodes have a positive initial attractiveness $(a > 0)$. If $k_0 = 1$, the target node in the Yule process is chosen proportional to the present number of links, $p_i = k_i / \sum_j k_j$. The probabilities for target nodes to have new links are constantly updated during the simulation.

(ii) **Frailty graph:**
Each node is assigned with a frailty $z_i$ at arrival. The frailty terms are generated from the distribution (18). Once a new node has entered, a waiting time is calculated from $-Log(r/z_i)$, where $r$ is a uniformly distributed random number, to identify the next time the node will make a new connection. The waiting times are measured in absolute time and stored in a sorted list $L = \{(N_j, t_j)..\}$.

As before, a new link is added with probability $p_{link}$. The origin of the link is chosen randomly among the nodes, and the first entry node in the list $L$ is selected as target node. A new waiting time is calculated for the target node and it is replaced accordingly in $L$. With probability $p_{node}$ a new node arrives and if $k_0 =1$, the target node will be the first node in the sorted list. The node is replaced in the list after a new waiting time is calculated.

## NUMERICAL SIMULATIONS

We simulate in parallel graphs derived from the Yule process (Yule graphs) and the randomized Poisson graphs (frailty graphs) with equal values of the parameters $k_0, a, m$.

In the early growth phase the frailty graph has a more heterogeneous appearance compared to the Yule graph of the same size (Fig. 1). This is natural since the variability is a build-in property of the nodes. In contrast, the variance of the linking rates increases with time in the Yule process. Given a Yule graph of size $N$ with scaling constant $q$, we apply (18) to generate a frailty graph with similar scaling behavior (Fig. 2).

The recommended graphical method, which was mentioned in the introduction, is employed to test for PA in the evolving networks (Fig. 3). The cumulative mean number of new links during a small time interval $\Delta t$ is plotted as function of the initial connectivity of the nodes on log-log axes. The mean numbers are group average among all nodes with identical link numbers at $\Delta t = 0$. A line has been added showing the expected linear preference slope.

Interestingly, the two graphs are quite similar, implying that the suggested method cannot distinguish a random process acting on nodes with heterogeneous rates from true preferential attachment. Hence, the assumption of PA has to be made *a priori*.

The different linking dynamics can be identified from the joint degree distribution (JDD) $P(k_1, k_2)$, portraying the interconnectedness between nodes. The distribution describes the probability that a randomly selected link has end points in nodes with connectivity $k_1$ and $k_2$. It has the definition:

$$P(k_1,k_2) = \frac{\alpha(k_1,k_2)m(k_1,k_2)}{2m} \qquad (20)$$

where $m$ is the number of links connecting $k_1, k_2$ types of nodes, and $\alpha(k_1,k_2) = 2$ for $k_1$ identical to $k_2$; otherwise $\alpha(k_1,k_2)$ is equal to unity.

The JDD of the frailty graph and the Yule graph give characteristic contours (Fig. 4A-B). The most frequent links in both graphs are edges connecting medium-degree nodes, producing an area with high frequency densities in the lower left corners. High-degree nodes are mostly connected to low-degree nodes (bottom right; top left).

We evaluate the link densities by plotting the difference $JDD_{frailty} - JDD_{Yule}$ (Fig. 4C). The frailty-based topology has the largest density of links connecting medium-degree nodes; the area with high density is extended in the Yule-based graph and stretched towards the axes. Thus, the probability for low- or medium-degree nodes to be coupled to nodes of similar degree is highest in the frailty graph (Fig. 4C, red areas), whereas the Yule graph has excess of links aligned at the axes, and at the radial front (Fig. 4C, blue areas). The latter effect is caused by the PA process, where nodes with high degree are identical to nodes with an early start, and hence, they have a larger probability for being linked together.

The graphs can be compared to their static random graphs $JDD(random)$, by simply performing a random rewiring of the links, while retaining the degree of each node (Fig. 4D). The plot is made for the frailty graph and shows that the dynamic topology is distinctly deficient links interconnecting medium-degree nodes compared to the static graph (yellow-blue). There is an excess of links connecting nodes of different degree (red). Hence, a static random graph does not serve as a good proxy for the dynamical network. The same finding applies to the Yule graph (data not shown).

Another way to visualize the structural difference is by examining the network core and fringe size as function of time. The $k$-core is identical to the sub-graph obtained from removal of all nodes of degree less than $k$.

Particularly, the graph coreness is $k_{max}$ for which the core is not empty. In the other end, the fringe is the set of loosely connected nodes in the graph, corresponding to nodes with minimum coreness $k_{min}$.

The Yule graph characteristically develops small, but condensed network cores (Fig 5). Hence, the Yule network has higher $k_{max}$ values, but the size of the network core is smaller. The difference in relative core size depends on the parameters, and is more pronounced for graphs with higher density ($m > 2$) (data not shown).

In contrast, the Yule graph has a larger group of loosely connected nodes, i.e. fringe size, compared to a frailty network of the same size (Fig. 5). This finding is robust and independent on link density. The difference in fringe size can be explained by the random entrance of low activity nodes in the frailty process, compared to the correlation between young age and low activity in the Yule process.

# CONCLUSION

We have provided a framework for simulating frailty random graphs with power law distributed degree distributions. Nodes are assigned with a random frailty variable, and the frailty distribution defines the variation in linkage rates among the nodes. Thus, the notion of frailty is closely related to the concept of fitness as a measure of relative reproductive success.

The important finding in this article is the existence of spurious preferential attachment (PA). Generative network models commonly involve PA mechanisms, but often no effort is made to actually confirm proportionate growth from network data. We have demonstrated that nodes in a randomized Poisson process and a Yule process exhibit similar tendencies for having new links when grouped by their previous connectivity. Hence, the conjecture of PA cannot be tested from simple graphical plots.

It should be emphasized that we do not endeavor to disregard PA as being an important process in evolution of real-world networks. Indeed, there are cases where the PA mechanism seems well grounded. The prime example is the well-studied World Wide Web network [2], where advanced search engines make direct use of previous linking rates to rank retrieved information. For most other types of networks, particularly those with a more limited natural scale, inherent heterogeneity among nodes is potentially of great importance and cannot be neglected.

The present models are highly simplistic and do not reproduce important structures like local clustering, which is observed in natural networks. However, the purpose here has not been to make realistic network models, but to provide focus to the problem of identifying driving mechanisms for large-scale structures in networks.

## *ACKNOWLEDGEMENTS*

B.F.de Blasio was supported by the Norwegian Research Council.

## *REFERENCES*

1. A. L. Barabasi and R. Albert, Science 286, 509 (1999).
2. Dorogovtsev SN and Mendes JFF, Evolution of Networks-From Biological Nets to the Internet and WWW (Oxford University Press, Oxford, 2003).
3. G. Yule, J Royal Stat Soc 88, 433 (1925).
4. M. Newman, Contemporary Physics 46, 323 (2005).
5. Krapivsky PL, Redner S, and Leyvraz F, Phys Rev Lett 85, 4629 (2000).
6. Dorogovtsev SN, Mendes JFF, and Samukhin AN, Phys Rev Lett 85, 4633 (2000).
7. Molloy M and Reed B, Rand Struct Alg 6, 161 (1995).
8. G. Caldarelli, A. Capocci, R. P. De Los, and M. A. Munoz, Phys. Rev. Lett. 89, 258702 (2002).
9. Merton RK, Science 159, 56 (1968).
10. Liljeros F et al., Nature 411, 907 (2001).
11. M. Taibleson, Am Soc Rev 39, 877 (1974).
12. Eggenberge F and Polya G, Zeitschrift für Angewandte Mathematik 3, 279 (1923).
13. Greenwood M and Yule GU, J Royal Stat Soc 83, 255 (1920).
14. Jeong H, Néda Z, and Barabasi L, Euro Phys Lett 61, 567 (2001).
15. Newman MEJ, Phys Rev E 64, 025102-025102-4 (2001).
16. B. de Blasio, Svensson Å, and F. Liljeros, PNAS (2007).
17. J. Willis, Age and Area Cambridge, 1922).
18. Simon HA, Biometrika 42, 425 (1955).
19. M. Abramowitz and I. Stegun, in Handbook of Mathematical Functions, Edited by National Bureau of Standards Washington DC, 1972), Chap. 13.
20. J. Abate and W. Whitt, Informs Journal on Computing 11, 394 (1999).
21. J. Abate and W. Whitt, Stoch Models 15, 517 (1999).
22. WolframResearch. http://functions.wolfram.com/HypergeometricFunctions/Hypergeometric2F1/06/01/02/01 . 2007.

23. A. Feller, in An Introduction to Probability Theory and Its Applications, Second Edition ed., Edited by I. John Wiley & Sons New York, 1968), Chap. 13.
24. Miller GA, Am J Physchol 70, 311 (1957).
25. Reed WJ and Hughes BD, Phys Rev E 66, 067103-067103-4 (2002).

# FIGURE LEGENDS

**Figure 1: Yule graphs and frailty graphs:**
Two examples of graphs generated with different model parameters
A) Frailty graph $(k_0 = 1; a = 0; m = 1)$ $n = 800$ nodes; B) Yule graph $(k_0 = 1; a = 0; m = 1)$ $n = 800$ nodes
C) Frailty graph $(k_0 = 1; a = 0; m = 4)$ $n = 400$ nodes; D) Yule graph $(k_0 = 1; a = 0; m = 4)$ $n = 400$ nodes. Plots were made using the freeware program *Pajek*.

**Figure 2: Cumulative degree distribution:**
The distribution of the Yule graph (circles) and the frailty graph (squares) is shown. Network size is $N = 50.000$, and model parameters are: $k_0 = 1$; $a = 0$; $m = 2$; $\theta = 0.9985$.

**Figure 3:**
**Estimation of PA**:
The cumulative mean number of links grouped by previous connectivity for a Yule graph (circles) and a frailty graph (squares). The linkage rate is tested on networks of size $N = 10^5$ in an interval of $\Delta N = 10^3$. Model parameters are: $k_0 = 1$; $a = 0$; $m = 2$; $\theta = 0.999$.

**Figure 4:**
**Joint degree distribution (JDD)**:
**A)** frailty graph and **B)** Yule graph; parameter values are: $k_0 = 1; a = 0; m = 4$; graph sizes $N = 30000$. **C)** The difference between the distribution $JDD_{frailty} - JDD_{Yule}$, parameter values as above. **D)** $JDD_{frailty} - JDD_{frailty}(random)$. Both networks are weakly dissortative with assortativity coefficients of $r = -0.052$ for the frailty graph, and $r = -0.0051$ for the Yule graph.

**Figure 5**:

**Coreness**:

The relative size of the network core ($k_{max}$) is shown for the Yule graph (circles) and frailty graph (triangles). The relative fringe size of the Yule graph (squares) and frailty graph (diamonds) is also shown.

For the Yule graph the corresponding values of the network core and mean coreness ($N = 50.000$) are: $k_{max} = 14$ and $<k_{core}> = 4.097$; the values of the frailty graph are: $k_{max} = 11$ and $<k_{core}> = 4.108$.

Model parameters $k_0 = 1; a = 0; m = 3..$

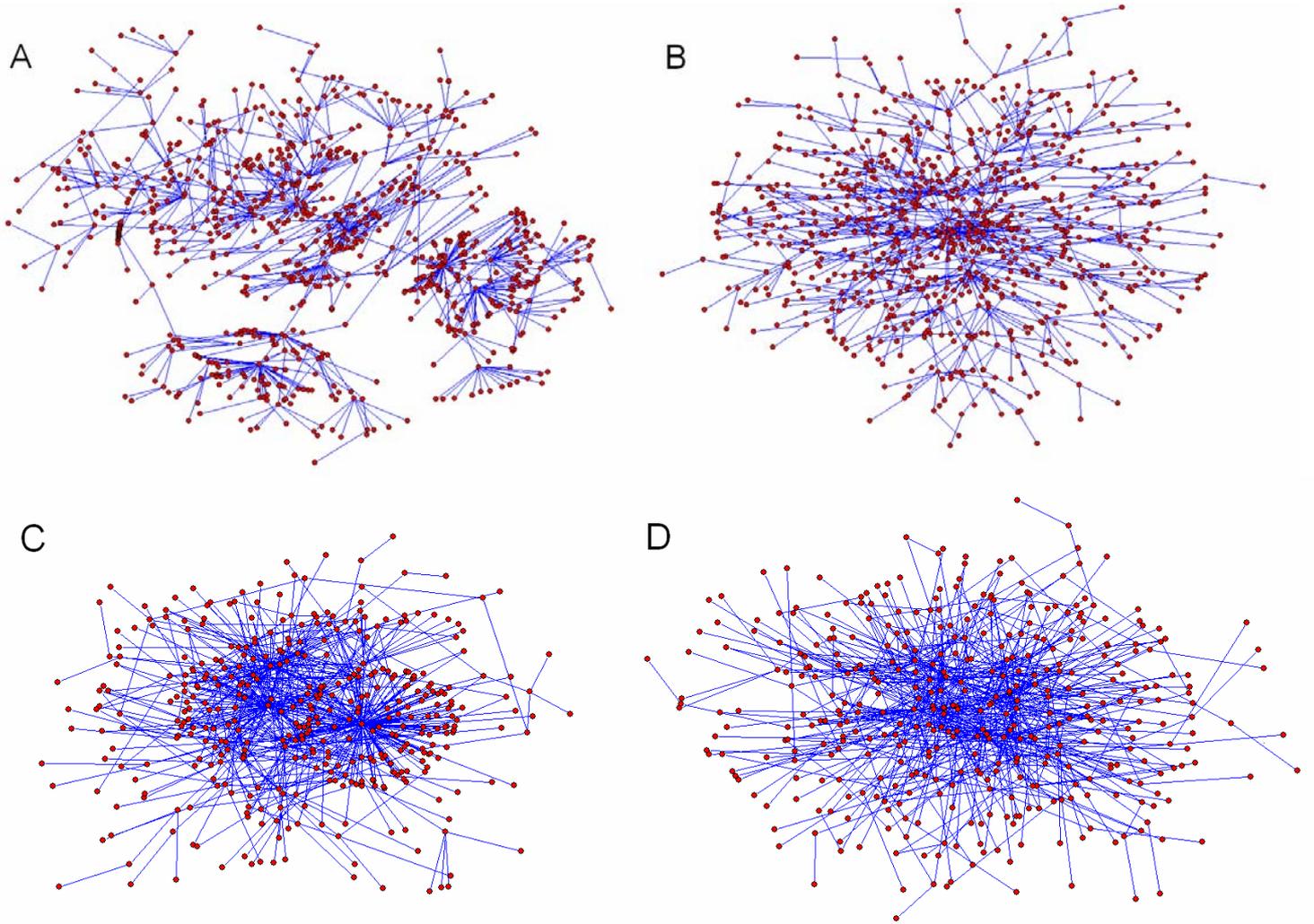

**FIGURE 1**

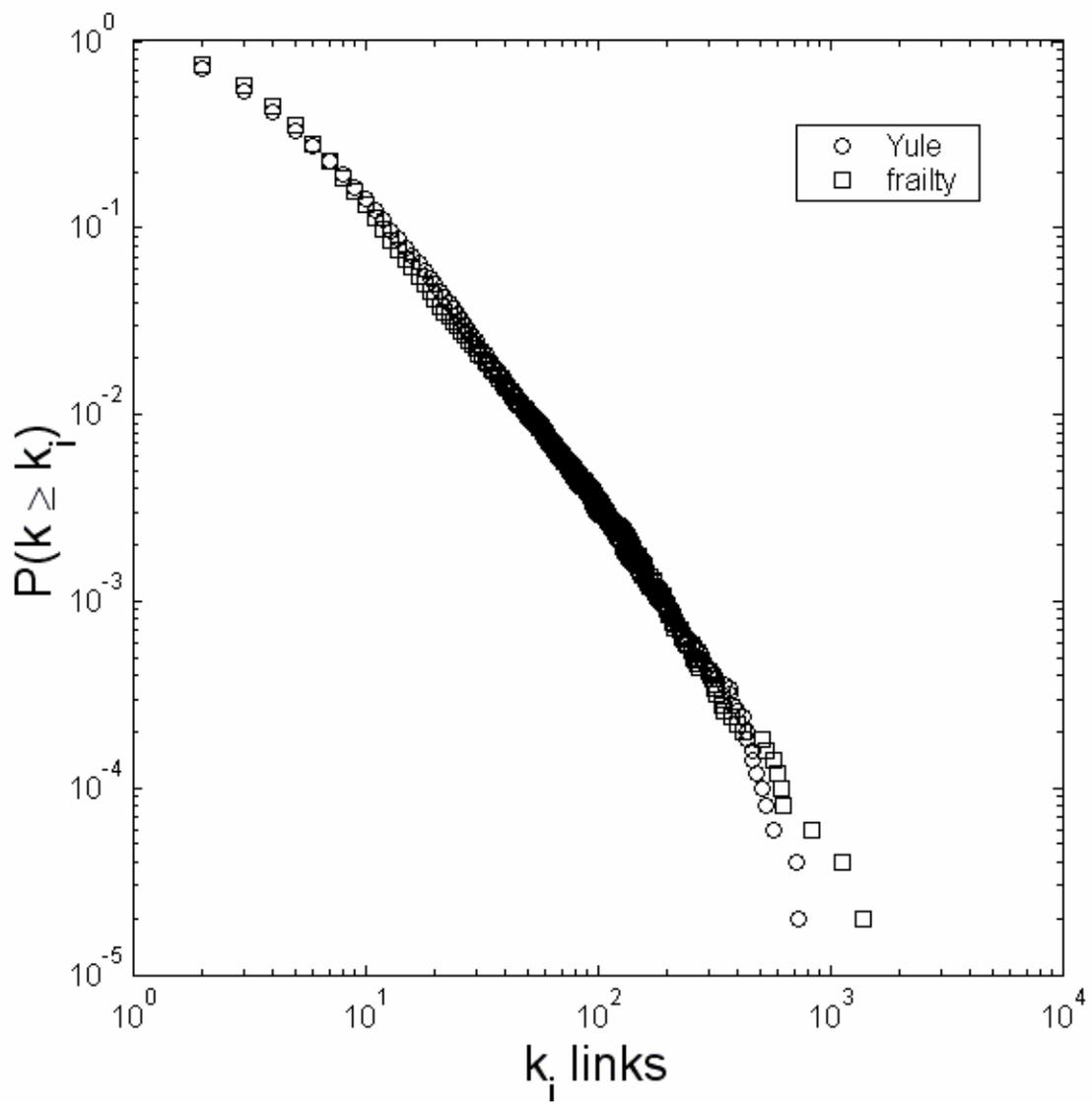

**FIGURE 2**

'

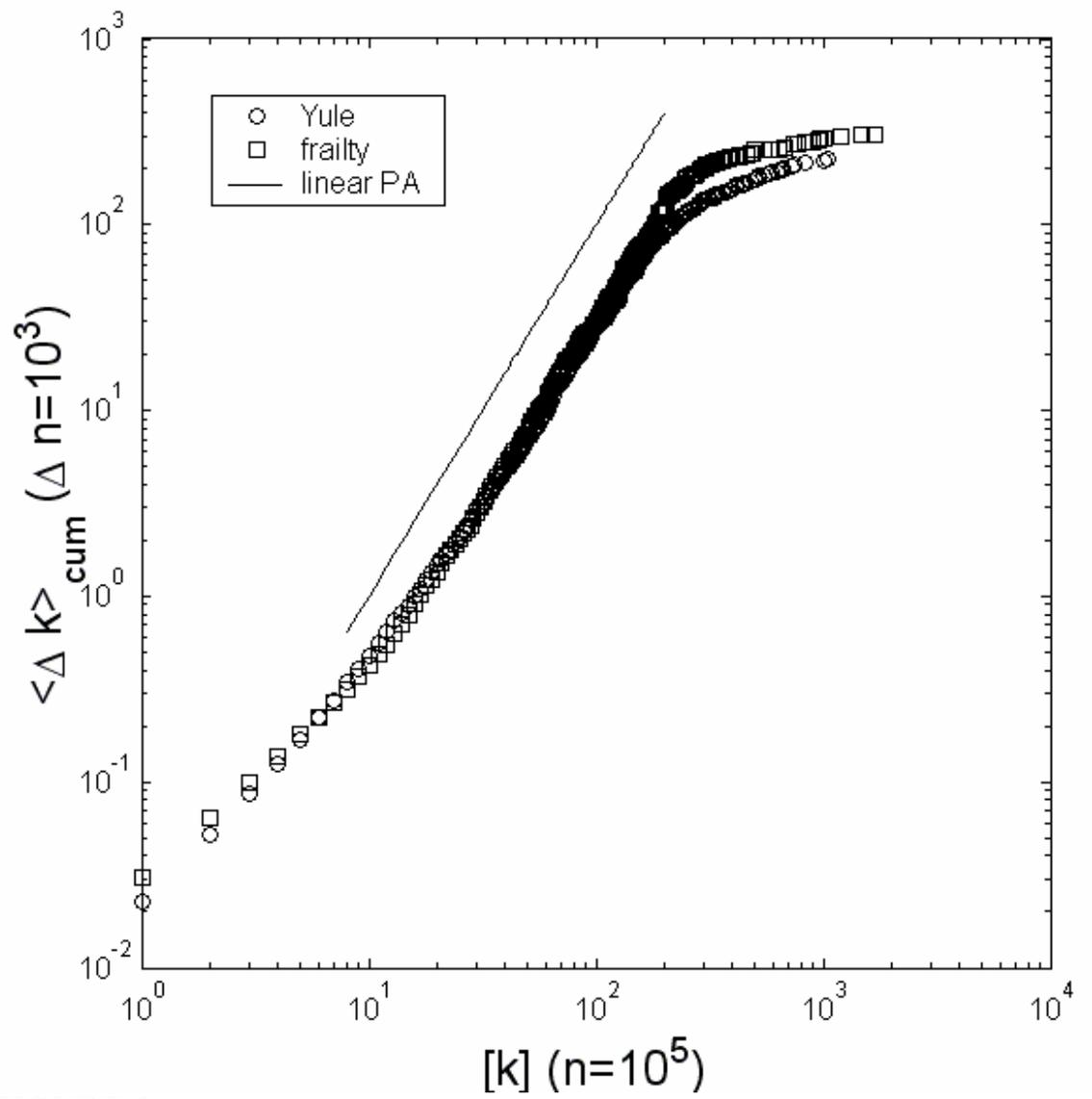

**FIGURE 3**

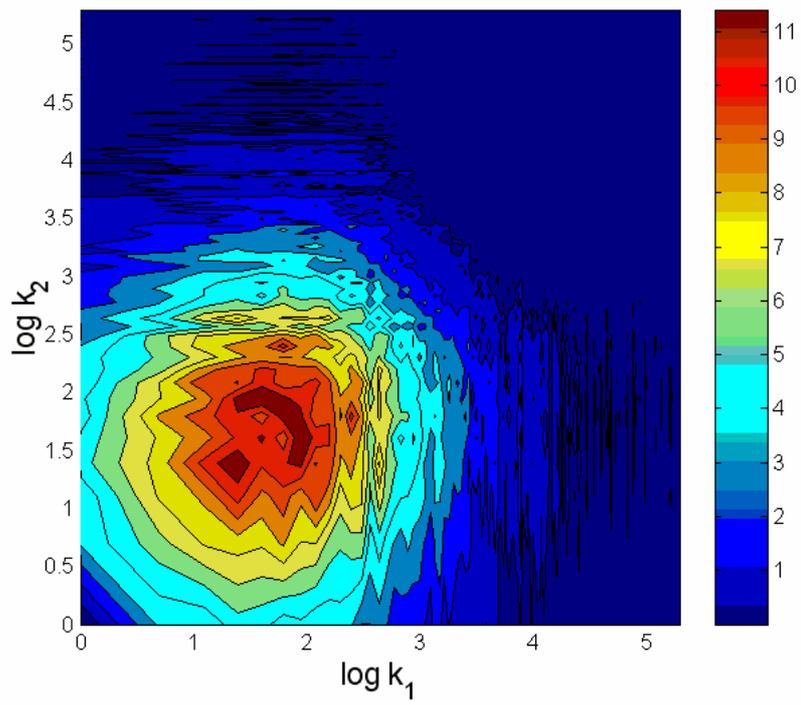

**FIGURE 4A**

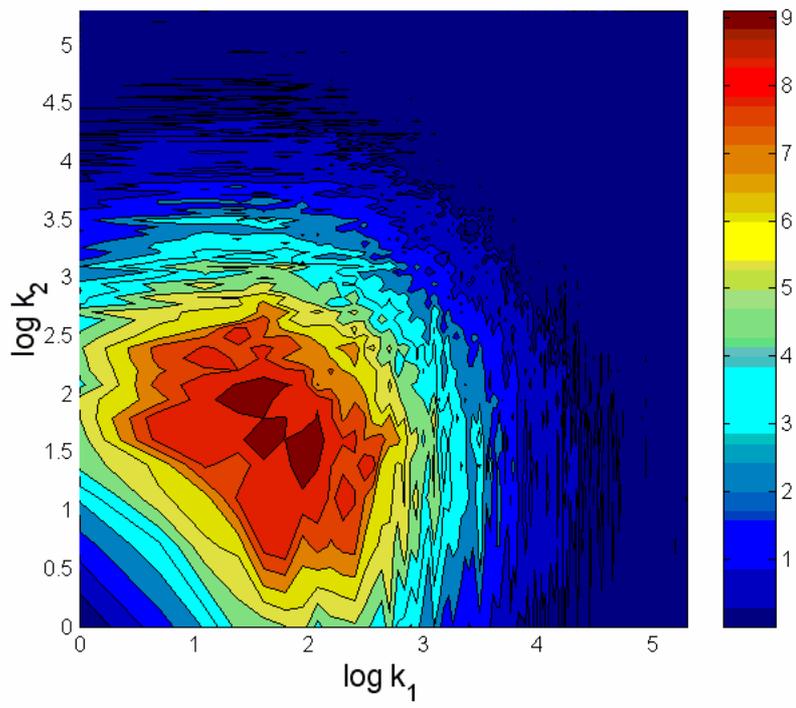

**FIGURE 4B**

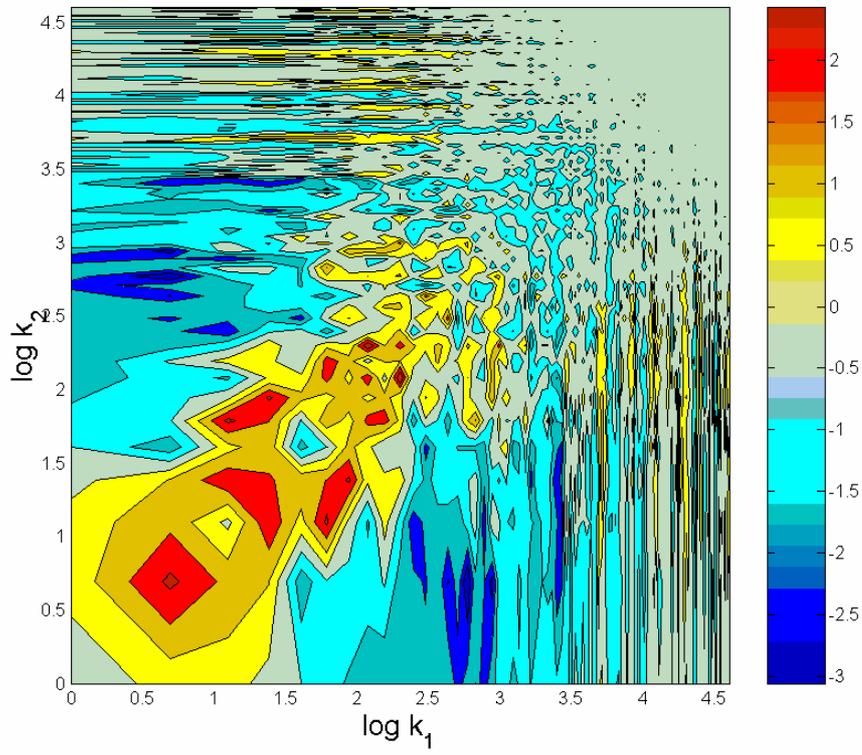

**FIGURE 4C**

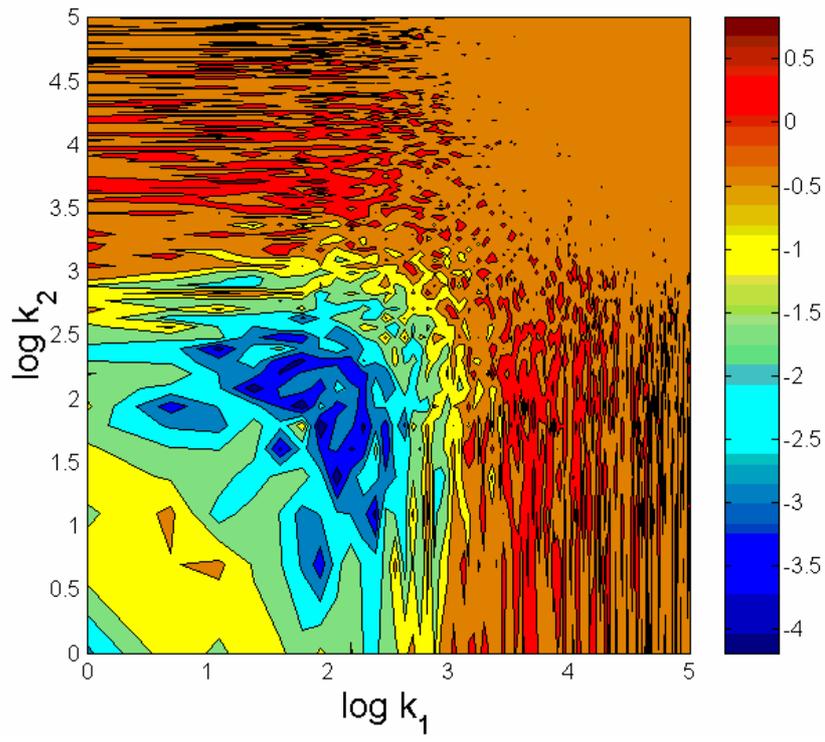

**FIGURE 4D**

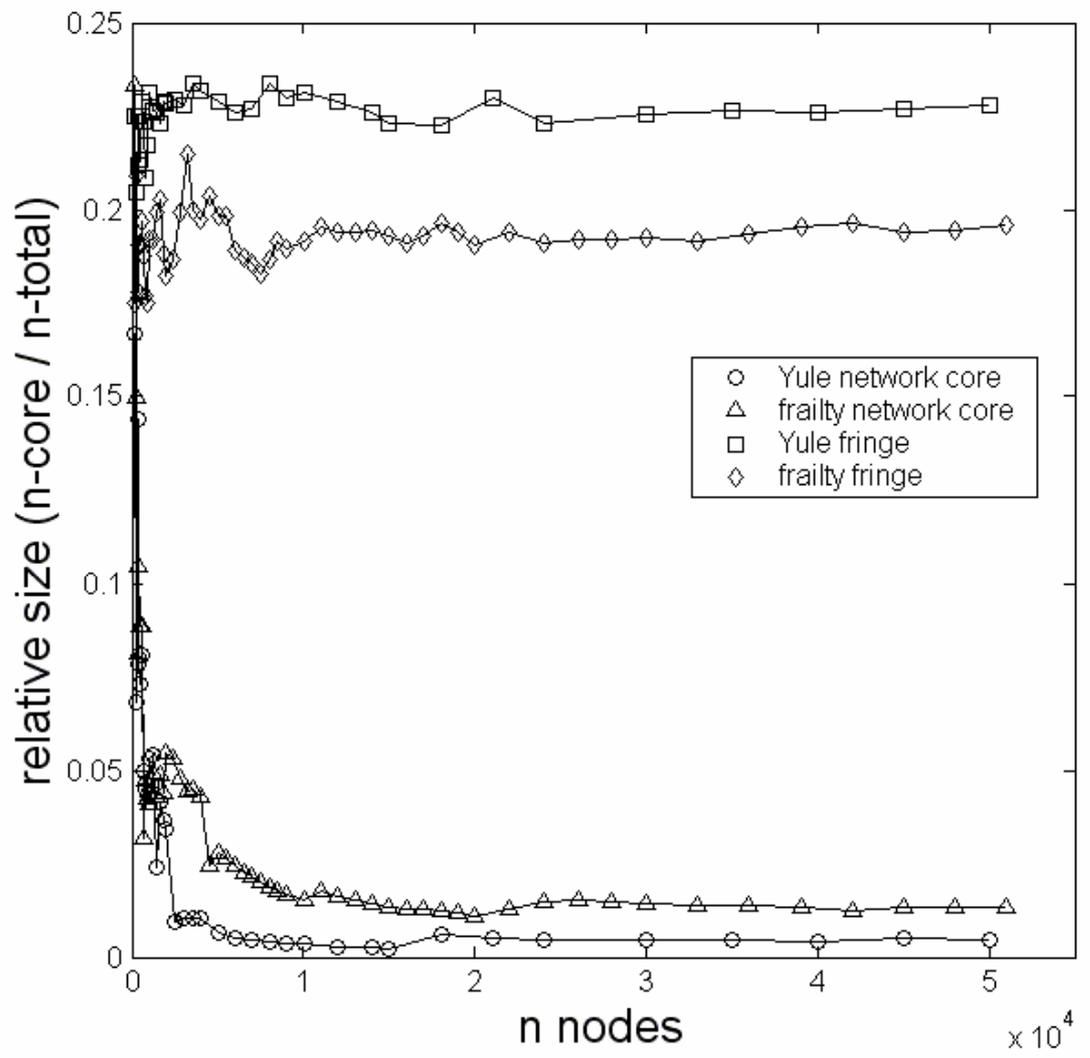

**FIGURE 5**